\documentclass[sigconf]{acmart} 
\usepackage[utf8]{inputenc}
\usepackage{hyperref}
\usepackage{graphicx}

\usepackage{soul}

\setcopyright{none} 
\renewcommand\footnotetextcopyrightpermission[1]{} 

\title{NASA Science Mission Directorate Knowledge Graph Discovery}

\copyrightyear{2023}
\acmYear{2023}
\setcopyright{licensedusgovmixed}\acmConference[WWW '23 Companion]{Companion Proceedings of the ACM Web Conference 2023}{April 30-May 4, 2023}{Austin, TX, USA}
\acmBooktitle{Companion Proceedings of the ACM Web Conference 2023 (WWW '23 Companion), April 30-May 4, 2023, Austin, TX, USA}
\acmPrice{15.00}
\acmDOI{10.1145/3543873.3587585}
\acmISBN{978-1-4503-9419-2/23/04}

\author{Roelien C. Timmer}
\affiliation{%
  \institution{University of New South Wales}
  \country{Australia}
}

\author{Fech Scen Khoo}
\affiliation{%
  \institution{University of Oldenburg}
  \country{Germany}
}

\author{Megan Mark}
\affiliation{%
  \institution{Florida Institute of Technology}
  \country{United States}
}

\author{Marcella Scoczynski Ribeiro Martins}
\affiliation{%
  \institution{Federal University of Technology}
  \country{Brazil}
}

\author{Anamaria Berea}
\affiliation{%
  \institution{George Mason University}
  \country{United States}
}

\author{Gregory Renard}
\affiliation{%
  \institution{The Applied AI Company (AAICO)}
  \country{United States}
}

\author{Kaylin Bugbee}
\affiliation{%
  \institution{NASA Marshall Space Flight Center}
  \country{United States}
}

\begin{document}

\begin{abstract}
The size of the National Aeronautics and Space Administration (NASA) Science Mission Directorate (SMD) is growing exponentially, allowing researchers to make discoveries. However, making discoveries is challenging and time-consuming due to the size of the data catalogs, and as many concepts and data are indirectly connected. This paper proposes a pipeline to generate knowledge graphs (KGs) representing different NASA SMD domains. These KGs can be used as the basis for dataset search engines, saving researchers time and supporting them in finding new connections. We collected textual data and used several modern natural language processing (NLP) methods to create the nodes and the edges of the KGs. We explore the cross-domain connections, discuss our challenges, and provide future directions to inspire researchers working on similar challenges.
\end{abstract}


\noindent \textbf{NASA Science Mission Directorate Knowledge Graph Discovery}\\

\noindent   Roelien C. Timmer, Fech Scen Khoo, Megan Mark, Marcella
Scoczynski Ribeiro Martins, Anamaria Berea, Gregory Renard, Kaylin Bugbee\\

This is a \textbf{camera-ready version} of the article accepted for publication at the \emph{3rd International Workshop on Scientific Knowledge: Representation, Discovery, and Assessment Spring 2023} co-located with the \emph{Web Conference 2023}.\\

Please cite this pre-print version as follows.\\

\noindent {\fontfamily{qcr}\selectfont @inproceedings\{timmer2023nasa,\\
\noindent author = \{Timmer, Roelien C. and Khoo, Fech Scen and Mark, Megan and Ribeiro Martins, Marcella Scoczynski and Berea, Anamaria and Renard, Gregory and Bugbee, Kaylin\},\\
title = \{NASA Science Mission Directorate Knowledge Graph Discovery\},\\
year = \{2023\},\\
maintitle =  \{The Web Conference 2023\},\\
booktitle = \{3rd International Workshop on Scientific Knowledge: Representation, Discovery, and Assessment Spring 2023\},\\
pages = \{(to appear)\},\\
location = \{Austin, Texas, USA\},\\       
\}

}
\newpage

\maketitle

\pagestyle{plain} 

\section*{Keywords}
Knowledge Graphs (KGs), National Aeronautics and Space Administration (NASA) Science Mission Directorate (SMD), Named-entity recognition (NER), Knowledge Representation, Natural Language Processing (NLP)

\section{Introduction}

As the interest in space exploration has increased, along with digitalization, the number of datasets in NASA's data catalog is growing exponentially. By 2023, the National Aeronautics and Space Administration (NASA) Science Mission Directorate (SMD) foresees the generation of at least 400 petabytes per year and expects this number to increase as new models are run, and new missions are launched~\cite{nasa2019strategy}. NASA has expressed the need for a search engine that enables scientists to search across the five NASA SMD scientific pursuits and allows them to discover new connections~\cite{bugbee2022selecting, duerr2021developing}. 

Search engines are often built with KGs as their backbone. KGs are a popular method to structure big and complex information. KGs organize information in a graph structure. There is no fixed definition of a KG, but in general, the nodes represent entities, and the edges represent the type of relationships~\cite{Ehrlinger2016TowardsAD}. KGs make it easier to digest complex information and make it easier to find connections between different concepts.

There are initiatives specifically for creating KGs for specific NASA SMD domains, such as the Heliophysics KNOWledge Network (Helio-KNOW) project~\cite{mcgranaghannasa}. Helio-KNOW is a community-build collection of software and systems to organize Heliophysics information\footnote{Link to Helio-KNOW project Github: \url{https://github.com/rmcgranaghan/Helio-KNOW}}. However, there are no KGs for all five of the NASA SMD domains. In addition, research is needed to connect the KGs of the five different NASA SMD scientific pursuits: (i) Heliophysics, (ii) Astrophysics, (iii) Planetary Science, (iv) Earth Science, and (v) Biological \& Physical Science. There has been an increased interest in multidisciplinary research between these different domains. For example, heliophysicists and astrophysicists can benefit from studying the high-energy solar flaring activity together to understand stellar flares~\cite{chen2022bridging}. Before merging the KGs graphs with, for example, entity alignment~\cite{trisedya2019entity,wu2019entity}, there is a need for a better understanding of the domain overlap. 

In this paper, we outline how we approached building KGs for the NASA SMD, of which the main steps are: (i) the collection of sufficient relevant textual data to train the NLP models, (ii) the selection of NLP algorithms to create the nodes and the edges of the KGs, (iii) exploring the overlap of the domains, (iv) validation, and, (v) visualization. While a significant contribution of this paper is providing a pipeline for generating KGs for the NASA SMD domains, our discussion of the challenges and recommendations for future work is valuable to other researchers. We also show our initial results of analyzing the overlap of the NASA SMD scientific pursuits.

This paper is structured as follows. We first outline related literature and work in Section~\ref{sec:related_work}. In Section~\ref{sec:methodology}, we propose a pipeline to generate KGs for the different NASA SMD domains. We present initial results in Section~\ref{sec:results}, including an overview of the data gathered and a cross-domain analysis. Finally, we have an extensive discussion of the challenges we faced and suggestions for future work in Section~\ref{sec:chal_fut} followed by a concise conclusion in Section~\ref{sec:conclusion}.

\section{Related Work}\label{sec:related_work}

In this section, we describe the main relevant existing research related to this work. We introduce the main concepts of KGs (Section~\ref{sec:related_work_KG}) and discuss methods used for scientific information processing (Section~\ref{sec:related_work_scientific}).

\subsection{Knowledge Graphs}\label{sec:related_work_KG}
The goal of KGs is to organize information in a graph structure. KGs have different definitions and forms, but in general, the nodes of a KG represent entities, i.e., concepts, objects, situations, or events, and the edges represent the relationship between the entities, which can both be qualitative and quantitative~\cite{Ehrlinger2016TowardsAD}. KGs gained popularity when big tech companies, such as Google, Microsoft, Facebook, and Yahoo!, created their own KGs and used them as the core of their semantic search engines. The three most well-known KGs are DBPedia~\cite{auer2007dbpedia}, Wikidata~\cite{vrandevcic2014wikidata}, and Google KG~\cite{singhal2012introducing}. 

\subsection{Scientific Information Processing}\label{sec:related_work_scientific}

In science, there are large volumes of complex information. KGs are popular~\cite{auer2018towards, luan2018multi} to structure all this information. One approach to generating scientific KGs is extracting entities from scientific documents~\cite{hou2021tdmsci, zong2021proceedings, gupta2011analyzing}. There are multiple named-entity recognition (NER) algorithms to extract scientific entities, such as AstroBERT~\cite{grezes2021building}, and SciBERT~\cite{beltagy2019scibert}. These scientific entities can serve as the nodes of the KGs. AstroBERT and SciBERT are based on Google's Bidirectional Encoder Representations from Transformers (BERT)~\cite{devlin2018bert}.

\section{Methodology}\label{sec:methodology}

\begin{figure*}[htbp]
    \centering
    \includegraphics[width=\textwidth]{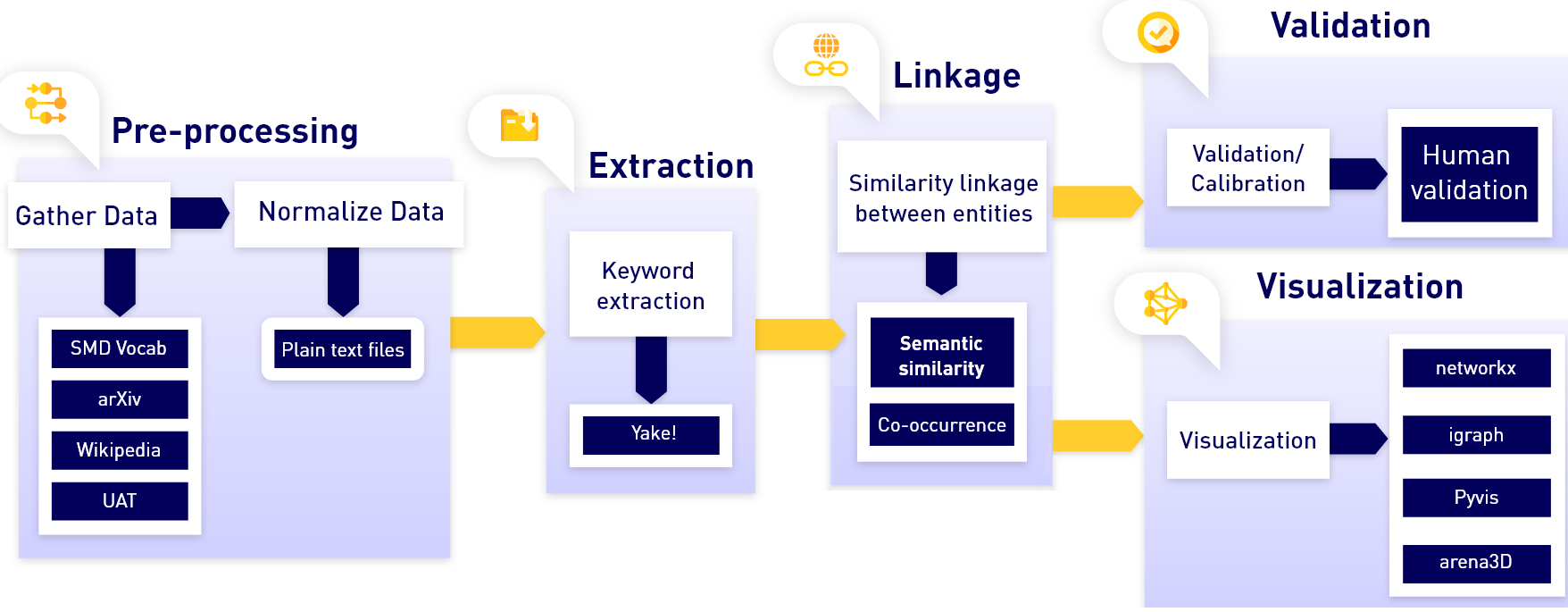}
    \caption{Proposed Pipeline for the Generation of Knowledge Graphs for the NASA SMD.}
    \label{fig:pipeline}
\end{figure*}

We propose a pipeline to generate KGs for the five NASA SMD domains. The pipeline consists of five major components, pre-processing (Section~\ref{sec:methodology_preprocessing}), entity extraction (Section~\ref{sec:methodology_extraction}), entity linkage (Section~\ref{sec:methodology_linkage}),  validation (Section~\ref{sec:methodology_validation}), and visualization (Section~\ref{sec:methodology_visualization}),  which are visualized in Figure~\ref{fig:pipeline}. We describe all five components below.

\subsection{Pre-processing}\label{sec:methodology_preprocessing}

We gather textual data from different sources. We need a high amount of textual data as this will be the input of the NLP models. The NASA SMD provided a list with definitions from the five different Scientific Pursuits \footnote{Example of the definition of Coronal Mass Ejection: \emph{An eruption in the outer solar atmosphere that sends billions of tons of magnetized plasma clouds into interplanetary space. When traveling at high speeds these ejections create shocks in the solar wind. Earth-intercept of a CME is often followed by a geomagnetic storm.}}. These definitions are a valuable source because they are created by NASA and used internally by researchers. We also downloaded terms from the Unified Astronomy Thesaurus (UAT), which is a  community-supported, open-source project from the American Astronomical Society and builds upon the International Astronomical Union (IAU) Thesaurus\footnote{Link to the UAT glossary: \url{https://astrothesaurus.org}\\Link to the UAT glossary in CSV format: \url{https://github.com/astrothesaurus/UAT/blob/master/UAT.csv}}\cite{accomazzi2014unified}. The UAT consists of astronomical concepts and their inter-relationships. The strength of the UAT is that anyone in the community can contribute by providing additions, refinements, and revisions; therefore, the UAT's quality and quantity are constantly being improved. To have sufficient data to train NLP models, we want to gather more textual data. We opt to web scrape Wikipedia summaries and arXiv abstracts based on the UAT terms. We web scrape Wikipedia because  of the broad coverage of relevant concepts and arXiv abstracts because they contain a lot of relevant technical terms.   In Section~\ref{sec:resuts_data}, we provide more details about the web scrape method we applied and how much textual data we gathered from all four sources.

The textual data is stored in different formats, such as Portable Document Format (PDF), Comma-Separated Values (CSV), Computable Document Format (CDF), and HyperText Markup Language (HTML), and therefore needs to be normalized to ensure that all the texts have the same format. We tokenize and lemmatize all the texts. With tokenization, the texts are divided into tokens, and with lemmatization, all the words of the texts are converted to their roots. We also remove stopwords and identify common bigrams and trigrams.

\subsection{Extraction}\label{sec:methodology_extraction}
To generate the nodes of the KGs, we extract the keywords from the texts. We use Yet Another Keyword Extractor, better known as Yake!\footnote{Link to Yake!: \url{https://github.com/LIAAD/yake}}~\cite{campos2020yake}. Yake! is an unsupervised automatic keyword algorithm that returns a  list of keywords and a corresponding quantitative score indicating the relevance of each keyword\footnote{The Yake! keyword is more relevant with a lower score.}. Examples of keywords extracted with Yake! from the UAT are \emph{Tauri stars}, \emph{Kerr black holes}, \emph{Classical novae}, and \emph{Hubble Space Telescope} \footnote{More examples of keywords extracted with Yake! can be found in~\cite{mark2022knowledge}.}. In Section~\ref{sec:discussion_nodes}, we discuss other methods that we tried and tested to extract entities to generate the nodes of the KGs.
 
\subsection{Linkage}\label{sec:methodology_linkage}
To link the nodes in the KGs, we considered two different techniques. First, count the frequency of the keywords co-occurring in a sentence as subjects and objects. The higher the frequency, the stronger the edge between the two nodes. Second, we quantify the strength of the edges by calculating the semantic similarity between NEs with the help of large pre-trained NLP models. For example, we finetune a model of the BERT model based on all the textual data we collected and calculate the semantic similarity between the nodes~\cite{devlin2018bert}.


\subsection{Validation}\label{sec:methodology_validation}
Validating the KGs is a big challenge as the nodes of the KGs mainly contain technical terms that only experts are fully familiar with. Examples of the nodes are: \emph{Fra Mauro}, \emph{Galilean Moons}, \emph{Chirality}, \emph{YORP Effect}, \emph{Nectarian} and \emph{Caldera}. Therefore, to validate the quality of the KGs, we validated the quality ourselves and invited domain experts to provide feedback. In Section~\ref{sec:discussion_validation}, we discuss alternative validation techniques.

\subsection{Visualization}\label{sec:methodology_visualization}
To gain a deeper understanding of the KGs, we create visualizations with NetworkX\footnote{Link to NetworkX: \url{https://networkx.org/}}~\cite{hagberg2008exploring}, iGraph\footnote{Link to iGraph: \url{https://igraph.org/}}\cite{csardi2006igraph}, Pyvis\footnote{Link to Pyvis: \url{https://pyvis.readthedocs.io/en/latest/}}\cite{perrone2020network}, and arena3D\footnote{Link to Arena3D: \url{https://arena3d.org/}}\cite{pavlopoulos2008arena3d}. All these four software packages have their own benefits and shortcomings. These visualizations assist us in validating the quality of the KGs but can also assist researchers make discoveries. 

\section{Initial Results}\label{sec:results}

In this section, we would like to discuss some initial results. This includes describing the type of data we collected (Section~\ref{sec:resuts_data}) and the results of exploring the overlap between some of NASA SMD's scientific pursuits (Section~\ref{sec:results_overlappingdomains}).

\subsection{Data Collection}\label{sec:resuts_data}

We collected data from four different sources: (i) SMD, (ii) arXiv, (iii) Wikipedia, and (iv) UAT. We used a definition list provided by the NASA SMD consisting of more than three million definitions and the UAT consisting of 2,826 terms. We web scraped relevant Wikipedia summaries and arXiv abstracts based on these UAT terms.

We used a collection of all the arXiv abstracts up to the end of 2021, a total of approximately two million abstracts \footnote{Link to the arXiv abstracts: \url{https://huggingface.co/datasets/gfissore/arxiv-abstracts-2021/blob/main/README.md}}. We filtered for abstracts that are related to the UAT terms. Specifically, for each abstract of this collection, we extracted the main 20 keywords with Yake! and selected the abstracts for which the semantic similarity between the 20 keywords and the UAT terms surpasses a certain threshold. We opted for 20 keywords as this was a balanced trade-off between the quantity and the quality of keywords. While we wanted to collect a high number of keywords, we also needed the keywords to be relevant.  In total, we ended up with 168,084 relevant arXiv abstracts. We followed a similar procedure for selecting relevant Wikipedia abstracts, selecting 695 summaries\footnote{We used the \emph{wikipedia} module of Python to parse summaries from Wikipedia. Link to the \emph{wikipedia} module: \url{https://pypi.org/project/wikipedia/}}. Table~\ref{tab:data} shows the main statistics, source, size, and type of all the textual data we gathered.

\begin{table}[]
\caption{A Summary of all the Textual Data Collected.}
\label{tab:data}
\begin{tabular}{cll}
    \toprule
    Source      & Size & Type \\
    \midrule
    Science Mission Directorate & 3,096,448 & definitions \\
    arXiv           & 168,084 & abstracts \\
    Wikipedia       & 695 & summaries \\
    Unified Astronomy Thesaurus & 2,826 & terms \\
    \bottomrule
\end{tabular}
\end{table}

\subsection{Overlapping Domains}\label{sec:results_overlappingdomains}

In the future, we want to merge the KGs of the different NASA SMD domains with entity alignment~\cite{trisedya2019entity,wu2019entity}. Before merging the KGs, we want to better understand how the different domains overlap. We experimented with k-means, an unsupervised clustering algorithm where all observations are partitioned into k clusters where each observation belongs to the cluster with nearest mean\footnote{For k-means, we use the \emph{scikit-learn}, also known as \emph{sklearn}, a module of Python. Link to \emph{scikit-learn} module: \url{https://scikit-learn.org/}}\cite{lloyd1982least}. 

We applied k-means to cluster a random selection of 1,500 Planetary Science and 1,500 Heliophysics arXiv abstracts without providing domain labels. We set k, the number of clusters, equal to two, as we were interested in whether the k-means algorithm can broadly distinguish between the two domains. K-means managed to classify the abstracts with greater than 80\% accuracy. This high accuracy indicates a distinct difference between these two domains. Contrarily, we suspect the accuracy was not perfect due to a significant overlap in terminology between the domains. 


\section{Challenges and Future Directions}\label{sec:chal_fut}

Many challenges arose during the effort and may show to be a continuous challenge as the work continues to improve the KGs. In this section, we will discuss our challenges concerning validation (Section~\ref{sec:discussion_validation}), access to validated ontologies (Section~\ref{sec:discussion_ontologies}), computational restraints (Section~\ref{sec:discussion_computational}), generating KG nodes (Section~\ref{sec:discussion_nodes}), domain overlap (Section~\ref{sec:discussion_domainoverlap}), and visualization (Section~\ref{sec:discussion_visualization}).

\subsection{Validation}\label{sec:discussion_validation}

Validating KGs is our biggest challenge. We asked domain experts to assess the global quality of the KGs and provided feedback ourselves. Manually validating was time-consuming and due to time and resource constraints we only managed to validate parts of the KGs. The main feedback was that the KGs with the Yake! keywords were more appropriate. In Section~\ref{sec:discussion_nodes}, a deeper discussion about the quality of the nodes is given. In the future, we would like to include a human in the loop, so the quality of the KGs can continuously be improved. One option would be that users of the KGs can suggest changes, directly apply changes, or changes are made indirectly based on the users' behaviour~\cite{manzoor2022expanding}. In addition, we aim to implement automatic metrics such as Corroborative Fact Validation (COPAAL)~\cite{syed2019unsupervised}, Deep Fact Validation (DeFacto)~\cite{lehmann2012defacto}, or FactCheck~\cite{syed2018factcheck}. COPAAL is easy to implement as it implements path scoring solely based on the provided KG. COPAAL can be complemented by DeFacto and FactCheck as these two metrics validate the quality of the KG based on external knowledge resources. 

\subsection{Access to validated ontologies}\label{sec:discussion_ontologies}

We intended to use validated ontologies as the backbone of our KGs. Unfortunately, we did not find relevant existing validated ontologies. We considered building ontologies from scratch and integrating them across the different NASA SMD domains, but eventually, we did not create them due to time constraints. In the future, we want to build ontologies manually and embed them in the KGs.

\subsection{Computational Restrains}\label{sec:discussion_computational}
Creating the KGs and experimenting with the data is computationally expensive and time-consuming. The big data sets, in combination with the different NLP algorithms, result in complex and intense processes. Due to time limitations, we limited the data size for some experiments. For example, for our domain overlap experiment (see Section~\ref{sec:results_overlappingdomains}), we limited the number of abstracts to 3,000 and only ran these experiments for two out of the five NASA SMD scientific pursuits.

\subsection{Generating nodes}\label{sec:discussion_nodes}
To generate nodes for the KGs, we considered multiple methods. We started by extracting the subject, predicate, and object for each sentence of our dataset. Instead, we found that the nodes generated by Yake! were the most suitable. For example for a sentence from the NASA SMD definition list \emph{They are made of cooler solar material, or plasma, supported in the sun's atmosphere by magnetic fields.}, the former extract \emph{they} and \emph{magnetic  fields} as the subject and object while the latter extracts \emph{solar materials} and \emph{magnetic fields} as keywords. We do believe that for the generation of the nodes, there is still the opportunity for improvement as Yake! is not specifically designed to recognize terms related to the NASA SMD.  In the future, we aim to create NER algorithms similar to AstroBERT~\cite{grezes2021building}, and SciBERT~\cite{beltagy2019scibert}.
In line with these two NER algorithms, we want to fine-tune models from the BERT family~\cite{devlin2018bert}.

\subsection{Domain Overlap}\label{sec:discussion_domainoverlap}
To promote cross-domain research, we need to generate five KGs, one for each of NASA's SMD domains, and we need to merge them. To merge the KGs, we need to apply entity alignment~\cite{trisedya2019entity,wu2019entity}. Entity alignment is a technique where entities of different KGs are identified to be referring to the same real-world object~\cite{zeng2021comprehensive}.  Entity alignment is challenging, as we face many homonyms. Homonyms are words that share the same spelling but have different meanings~\cite{hurford2007semantics}. For example, the term \emph{storms} could point to the geomagnetic storms for solar physicists or ice storms on earth.

\subsection{Visualization}\label{sec:discussion_visualization}
The KGs we created were too complex to visualize. Clustering the nodes helped a little, but we had too many nodes to create a visually enticing image. We visualized KGs based on the data selections, but we found that these KGs were too limited. In addition, generating interactive KG required too many computational resources and led to slowly working interactive KGs. In the future, we would like to build a robust Application Programming Interface (API) for 3-D visualization of the knowledge graphs to assist researchers in finding new connections. 

\section{Conclusion}\label{sec:conclusion}
The size of the NASA SMD data catalog is growing exponentially due to digitalization and the launch of new models. The NASA SMD promotes cross-domain opportunities, while historically, the data management of these domains has been in silos. The advancement of NLP allows us to process large amounts of textual data and generate cross-domain discovery KGs. This work presents a pipeline to generate KGs for the NASA SMD domains. We explore the domain overlap and discuss how the KGs can be merged in the future. 

\section{Acknowledgements}

This work has been enabled by the Frontier Development Lab\footnote{Link to Frontier Development Lab website: \url{https://frontierdevelopmentlab.org/}}. FDL is a collaboration between SETI Institute and Trillium Technologies Inc., in partnership with NASA, Google Cloud, Intel, NVIDIA, and many other public and private partners. TThe material is based upon work supported by NASA under award No(s) NNX14AT27A. Any opinions, findings, conclusions, or recommendations expressed in this material are those of the authors and do not necessarily reflect the views of the National Aeronautics and Space Administration. The authors would like to thank the FDL organizers, the SETI institute, Trillium, the FDL partners and sponsors, and the reviewers for their constructive comments during the research sprint.

\bibliographystyle{ACM-Reference-Format}
\bibliography{smd_kg}

\end{document}